# DIRECT NUMERICAL SIMULATION OF FLUID FLOW IN A 5X5 SQUARE ROD BUNDLE USING NEK5000

**Adam Kraus, Elia Merzari**
The Pennsylvania State University
State College, PA, USA

**Thomas Norddine**
Ecole des Ponts ParisTech
Marne-la-Vallee, France

**Oana Marin**
Argonne National Laboratory
Lemont, IL, USA

**Sofiane Benhamadouche**
Electricite de France
Palaiseau, France

## ABSTRACT

*Rod bundle flows are commonplace in nuclear engineering, and are present in light water reactors (LWRs) as well as other more advanced concepts. Inhomogeneities in the bundle cross section can lead to complex flow phenomena, including varying local conditions of turbulence. Despite the decades of numerical and experimental investigations regarding this topic, and the importance of elucidating the physics of the flow field, to date there are few publicly available direct numerical simulations (DNS) of the flow in multiple-pin rod bundles. Thus a multiple-pin DNS study can provide significant value toward reaching a deeper understanding of the flow physics, as well as a reference simulation for development of various reduced-resolution analysis techniques. To this end, DNS of the flow in a square 5x5 rod bundle at Reynolds number of 19,000 has been performed using the highly-parallel spectral element code Nek5000. The geometrical dimensions were representative of typical LWR fuel designs. The DNS was designed using microscales estimated from an advanced Reynolds-Averaged Navier-Stokes (RANS) model. Characteristics of the velocity field, Reynolds stresses, and anisotropy are presented in detail for various regions of the bundle. The turbulent kinetic energy budget is also presented and discussed.*

## 1. INTRODUCTION

Flow in rod bundles is encountered in the majority of nuclear reactor concepts, and characterizing this flow is crucial for predicting the thermal hydraulic performance of the system. The flow in rod bundles is complex due in part to the inhomogeneities in the cross-section, which differentiates it from canonical flows such as channel flow. This can lead to localized effects that can be difficult to predict. Thus for decades, much experimental and numerical research has been devoted to this topic. The earliest investigations were largely experimental by necessity, with detailed numerical research often impractical due to limited computing capabilities. Later, "low-resolution" system and subchannel codes were developed, using the experimental data to establish correlations for pressure drop, heat transfer, and so forth. These families of codes are still used substantially today, particularly in transient analyses where more detailed approaches are impractical. A major drawback, though, is their rather limited fidelity in treating the physical laws governing fluid flow, and reliance on empirical correlations from known flow conditions somewhat restricts their use in more exploratory situations.

In recent decades, the use of computational fluid dynamics (CFD) for investigating thermo-fluid behavior in nuclear systems has steadily increased and will continue to do so as the availability of computing power increases. Some key advantages of CFD over experiments include generally smaller cost, greater control over boundary conditions, as well as easy and non-intrusive access to data and statistics from the flow that can be very difficult to measure. CFD also attempts to provide a higher fidelity to the actual flow physics than do the low-resolution codes.

The issue of turbulence modeling and resolution of the relevant scales creates a hierarchy of CFD approaches. These fall into three major classes. Most routine CFD work today is done using the Reynolds-Averaged Navier-Stokes (RANS) approach, where the turbulence is treated as time-averaged and is entirely modeled via additional algebraic or differential equations, with coefficients that are often empirically derived from experiments and canonical flow data. Large Eddy Simulation (LES) resolves the largest turbulence scales and models the smallest scales, which are generally more isotropic and more universal in behavior. This comes at increased computational cost, and so use of LES is generally limited to relatively small domains, low Re, or both, unless high-performance computing (HPC) is used. The last class is Direct Numerical Simulation (DNS), sometimes also referred to as a "numerical experiment." In this approach, all relevant flow spatial and temporal scales are resolved, and a DNS is thus presumed to provide a faithful representation of the Navier-Stokes solution for the given problem.

However, the extreme resolution requirements, which scale roughly as $Re^{9/4}$, have historically confined DNS to very low Reynolds numbers and very small geometries. For this reason,

DNS of the flow in multiple-pin rod bundles are scarce, and the available DNS have primarily been limited to single-pin, infinite-lattice approaches. Since wall effects are always important in rod bundles, this somewhat limits the practical applicability of those approaches.

In the work presented here, we perform a DNS study of a 5 by 5 square rod bundle representative of LWR fuel using the spectral element code Nek5000. We believe this work will be of value to the engineering and fluids communities for multiple reasons. The data from this DNS should provide a source of data to elucidate deeper understanding of the flow physics in rod bundles. To this end, we present and discuss the velocity, Reynolds stresses, anisotropy, and turbulent kinetic energy (TKE) budget in various areas of the domain. The DNS can also be used to benchmark and assess the reliability of various turbulence modeling approaches.

## 2. MATERIALS AND METHODS

### 2.1 Code and Numerical Methods

DNS as performed in this work solves the incompressible Navier-Stokes equations of a Newtonian fluid in the absence of other body or external forces:

$$\frac{\partial u_i}{\partial x_i} = 0 \qquad (1)$$

$$\frac{\partial u_i}{\partial t} + \frac{\partial}{\partial x_j}(u_i u_j) = -\frac{1}{\rho}\frac{\partial p}{\partial x_i} + \nu \frac{\partial^2 u_i}{\partial x_j \partial x_j} \qquad (2)$$

where Eq. (1) maintains mass continuity, Eq. (2) maintains momentum continuity, ρ is the density of the fluid and ν represents the kinematic viscosity. Implicit summation applies. In the current work, energy is not considered, and the density and viscosity are treated as constant.

The DNS was performed using the open-source, massively-parallel spectral-element code Nek5000 [1]. This employs the Spectral Element Method (SEM), a Galerkin-type method developed by Patera and others [2]. SEM displays the minimal numerical dispersion and dissipation of spectral methods while also featuring the geometric flexibility of finite element methods. This makes SEM an excellent choice for DNS and LES simulation of simple to complex geometries. Nek5000 in particular is also highly-parallel which is crucial in enabling these very large, high-fidelity simulations.

In SEM, the domain is discretized into E hexahedral elements and represents the solution as a tensor-product of $N^{th}$-order Lagrange polynomials based on Gauss-Lobatto-Legendre (GLL) nodal points. This leads to roughly $E(N+1)^3$ degrees of freedom per scalar field. The discrete Poisson equation for the pressure is solved using a variational multigrid GMRES method with local overlapping Schwarz methods for element-based smoothing at resolution N and ≈ N/2, coupled with a global coarse-grid problem based on linear elements [3]. Viscous terms are treated implicitly with second-order backward differentiation, while non-linear terms are treated by a third order extrapolation scheme. Nek5000 has been validated extensively for various rod bundle flows, including cases with spacer grids and wire wraps [4-7].

### 2.2 Computational Domain and Mesh

The geometry for the DNS is a 5 by 5 square array of cylinders surrounded by a square wall. The array pitch P is 1.326D, which is a typical value seen in Pressurized Water Reactor fuel. The outer wall-to-wall normal distance is 5P. The length of the domain was set at 3.158D, which was chosen based on review and on some sensitivity studies performed with an LES approach with Nek5000 [8,9].

DNS requires resolving from the integral scales all the way down to the Kolmogorov scales at minimum. The finite volume code STAR-CCM+ [10], with an advanced RANS approach using the Elliptic Blending Reynolds-Stress Model [11], was used to estimate the Taylor ($\lambda=\sqrt{15k\nu/\varepsilon}$) and Kolmogorov ($\eta=(\nu^3/\varepsilon)^{1/4}$) scales. These results were used to establish the DNS mesh sizing such that the distance between GLL points was smaller than $\min(\lambda/2,\eta)$. The mesh was also designed with a target dimensionless wall distance of $y+ = 0.5$.

The computational mesh consists of 8 387 008 curvilinear hexahedral elements and the polynomial order was set at 7, thus roughly 4.3 billion grid points have been used for the simulation. The elements were generated using the open source package Gmsh, and the element structure is shown in Fig. 1. The GLL points were generated internally in Nek5000. Fig. 2 demonstrates that the mesh resolution is below the Kolmogorov scale throughout the domain.

### 2.3 Flow Conditions and Simulation Procedures

The DNS was run non-dimensionally. The DNS had a Reynolds number of 19,000 based on the rod diameter. While the Reynolds number in prototypical geometries will be considerably higher, a value of 19,000 is a good compromise between the need of being representative (i.e., sufficiently turbulent) and computational cost, which scales super-linearly with the Reynolds number.

A CFL number of 0.5 was targeted with an adaptive timestepper before beginning averaging, i.e. before achieving stationary turbulence in the domain. Then a constant time step with mean CFL ~ 0.3 was used for time averaging. The results presented here represent time averages of roughly 15 flow-through times.

In addition to the time averaging, spatial averaging was used to hasten the process of obtaining converged statistical averages. The flow was averaged over the streamwise direction since it is a direction of homogeneity. Additionally, the invariance with 90° rotations and the diagonal symmetry were used, allowing for averages to be obtained on 1/8 of the cross-section. The results presented in the following sections refer to time- and space-averaged quantities as described above unless stated otherwise.

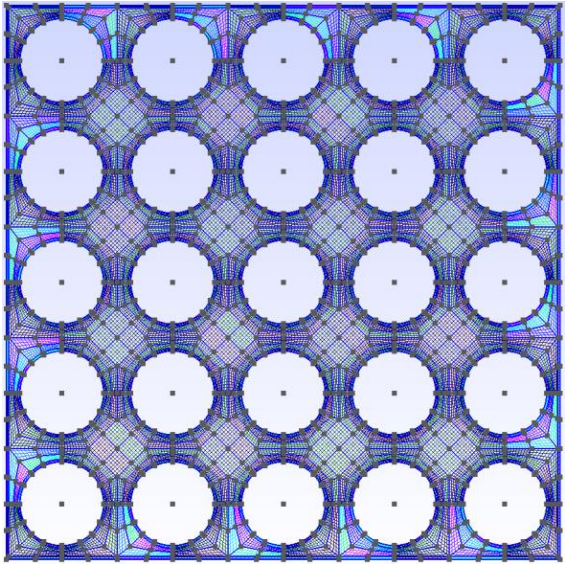

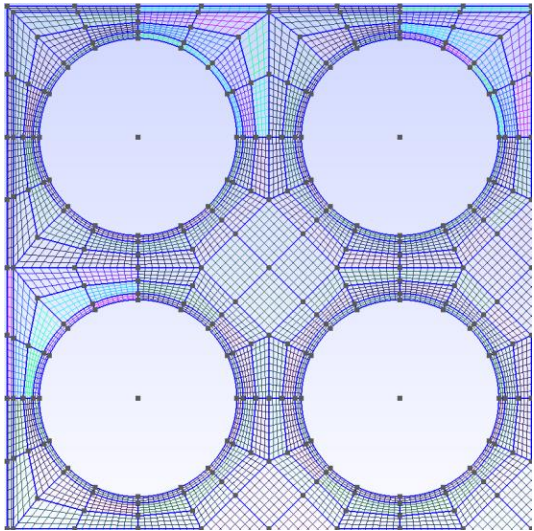

**FIGURE 1:** MESH ELEMENT STRUCTURE FROM GMSH USED IN NEK5000 RUNS (TOP) WITH ZOOMED VIEW NEAR THE CORNER SUBCHANNEL (BOTTOM)

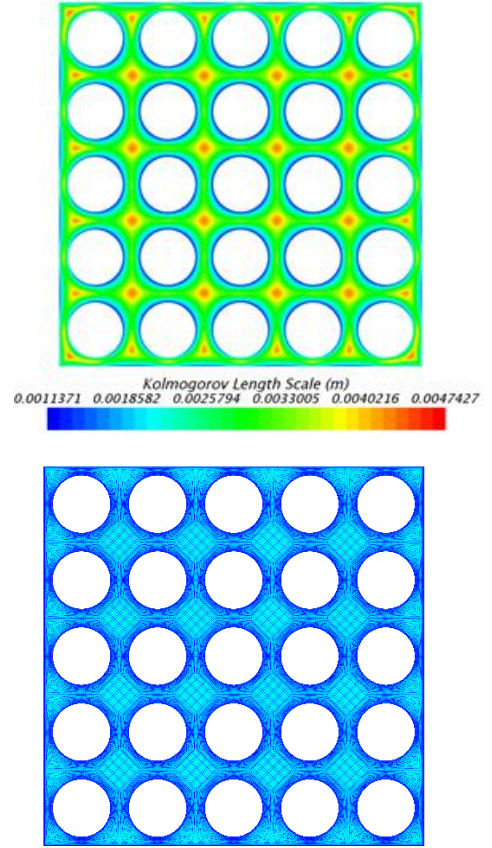

**FIGURE 2:** KOLMOGOROV LENGTH SCALE PREDICTED BY RANS (TOP) COMPARED TO DNS MESH SIZE (BOTTOM)

## 3. RESULTS AND DISCUSSION

### 3.1 Velocity Profiles

Fig. 3 provides the instantaneous and averaged streamwise velocity $w$ and demonstrates the primary features of the flow field. The wide range of simulated scales is easily seen. Generally speaking, the interior subchannels are rather similar in behavior, with a thin boundary layer and relatively uniform core region similar to channel flow. The edge subchannels, however, display more complex flow phenomena. Fig. 4 and Fig. 5 provide some instantaneous views of streamwise and wall-parallel velocity components, respectively, at the centerlines of the central subchannel as well as the edge subchannel. It is clear that there are significant differences in not only the relative velocity magnitudes in these regions, but also in the flow structures, which are impacted by the wall and the narrowed gaps.

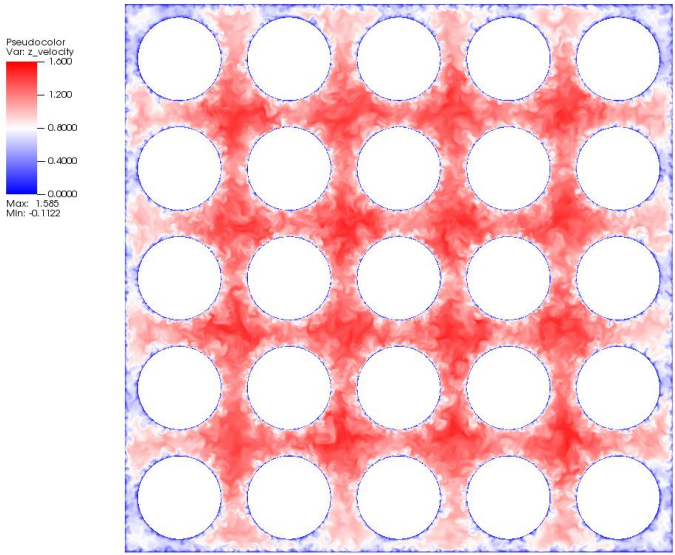

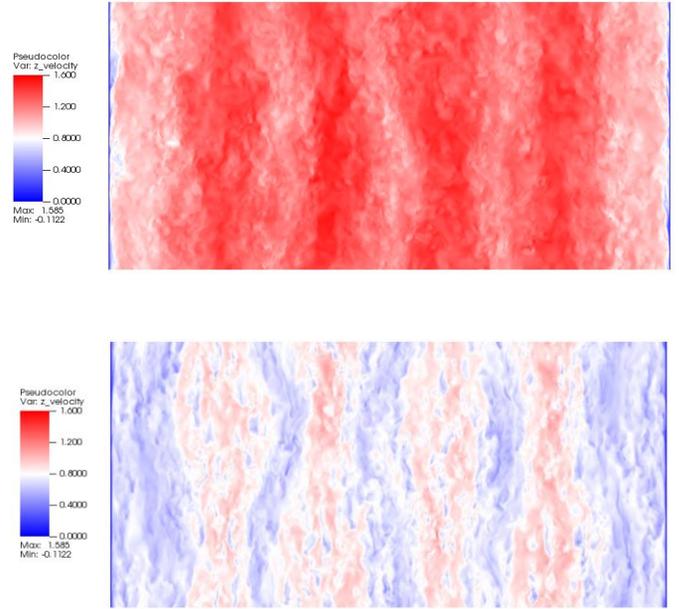

**FIGURE 4:** INSTANTANEOUS STREAMWISE VELOCITY IN THE CENTER SUBCHANNEL CENTERLINE (TOP) AND THE EDGE SUBCHANNEL CENTERLINE (BOTTOM)

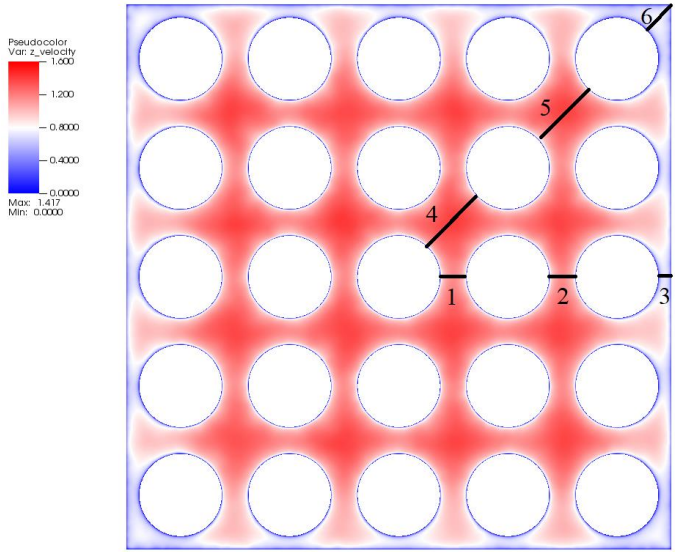

**FIGURE 3:** INSTANTANEOUS (TOP) AND TIME- AND STREAMWISE-AVERAGED (BOTTOM) STREAMWISE VELOCITY. BOTTOM PLOT ALSO DEMONSTRATES THE LINES USED FOR PLOTTING IN SUBSEQUENT FIGURES

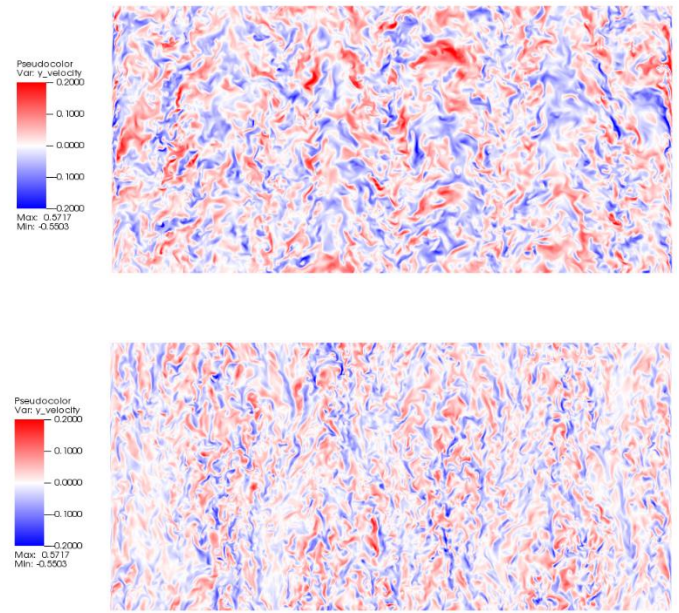

**FIGURE 5:** STREAMWISE WALL-NORMAL VELOCITY IN THE CENTER SUBCHANNEL CENTERLINE (TOP) AND THE EDGE SUBCHANNEL CENTERLINE (BOTTOM)

To compare the characteristics in a more quantitative manner, flow data were extracted along six different lines, as shown in Fig. 3. The first three lines are the "narrow" gaps proceeding outward from the center rod, such that line 3 is the edge subchannel. The last three lines are the "wide" gaps proceeding similarly outward along the domain diagonal, such that line 6 is the corner subchannel. In all plots, the coordinates begin nearest the domain center and proceed outward radially. The DNS results are also compared with LES results that were obtained on a roughly 1M element mesh in prior work [9].

Fig. 6 shows the streamwise velocity profiles along the sampling lines. Only the innermost half of each channel is shown. These are compared against the standard logarithmic law, given by

$$u^+ = \frac{1}{\kappa} \ln y^+ + C \qquad (3)$$

where constants of $\kappa = 0.41$ and $C = 5.2$ were employed here. The plots confirm that the flow profiles in the inner subchannels are indeed quite similar. The laminar sublayer and log region are both clearly demonstrated. The laminar region continues until $y^+ \approx 5$, and the log region begins at $y^+ \approx 25$, which is in line with values seen in canonical channel flow [12,13].

Flows in the edge and corner subchannels, i.e. Lines 3 and 6, show significantly different behavior. The local Re in these areas are too small relative to the shear for there to be a substantial log layer, and a clear deviation from the log law is seen. In the corner, the velocity gradient is already negative halfway along the channel width as well, and the flow profile along the channel is strongly asymmetric. These results suggest that "low-$y^+$" or adaptive wall function approaches would be beneficial for turbulence modeling in these regions.

**3.2 Reynolds Stresses**

The normal components of the Reynolds stress tensor (Fig. 7) are generally symmetric about the channel width, with only slight peaking on the centermost side. An exception is the corner subchannel which displays strong asymmetry. Note that all stress plots have been normalized to the peak stress, which is the streamwise normal stress $w'w'$. The majority of the components show strong "dual-peaking" near the two walls with a fast tailing off moving toward the center of the domain. This general behavior is true even for the corner subchannel, although the stress magnitude is less than half that seen in the interior subchannels. In these regions, the streamwise normal stress is always larger than the other two normal components.

The behavior of the normal stresses along Line 3 in the edge subchannel is unique and merits further discussion. While the streamwise component resembles that seen in the other channels, neither of the other two components are "double-peaked." Rather, these stresses have smooth profiles that have a central peak; in fact, the wall-parallel component is actually higher than $w'w'$ in this area. This behavior suggests there may be a gap vortex street in this region [14,15].

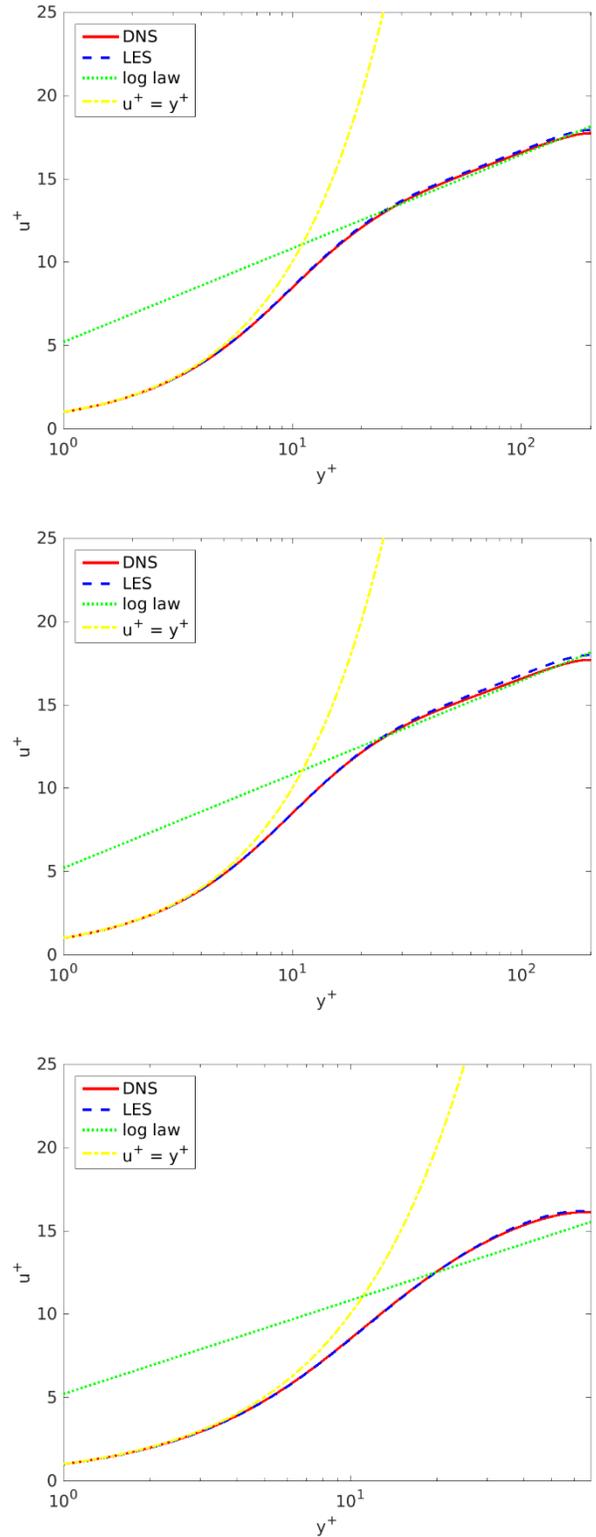

**FIGURE 6a:** NON-DIMENSIONAL VELOCITY PROFILES ACROSS THE CHANNEL HALF-WIDTHS FOR LINES 1 (TOP) THROUGH 3 (BOTTOM)

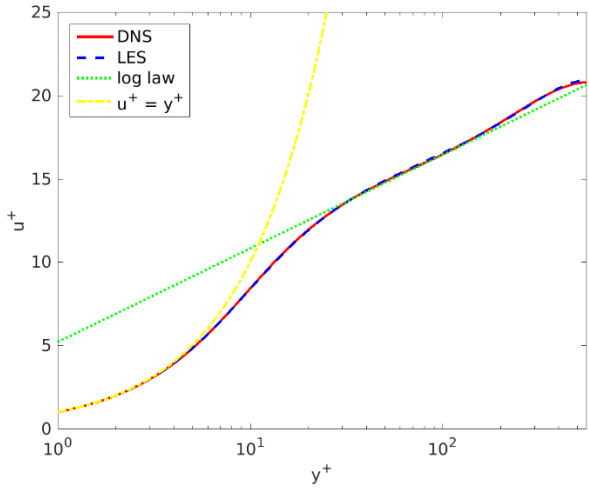
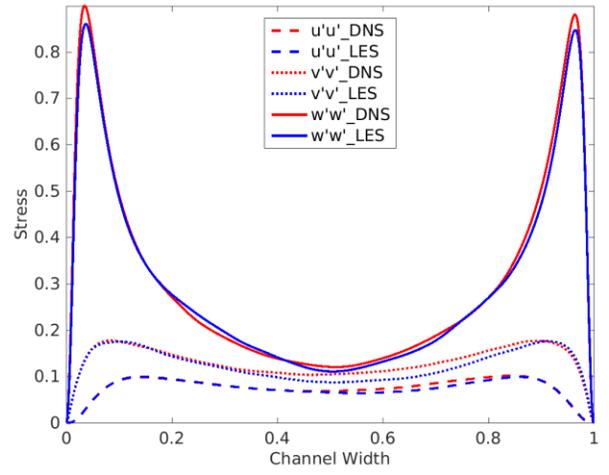
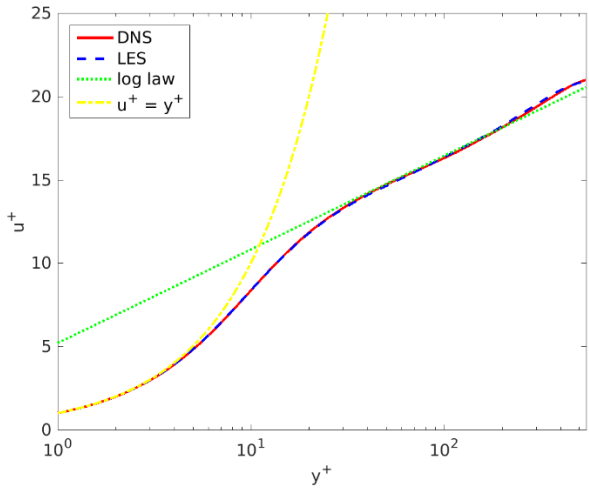
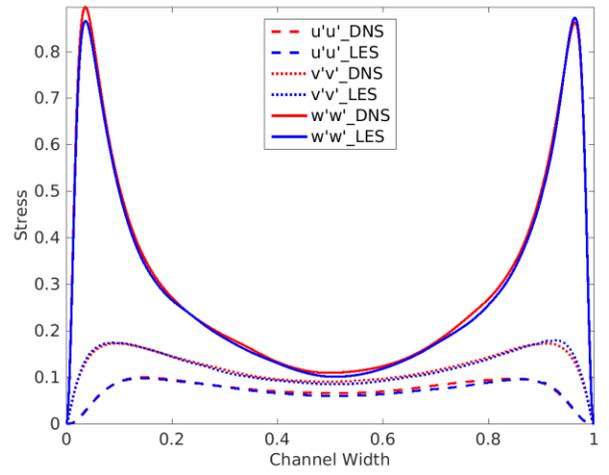
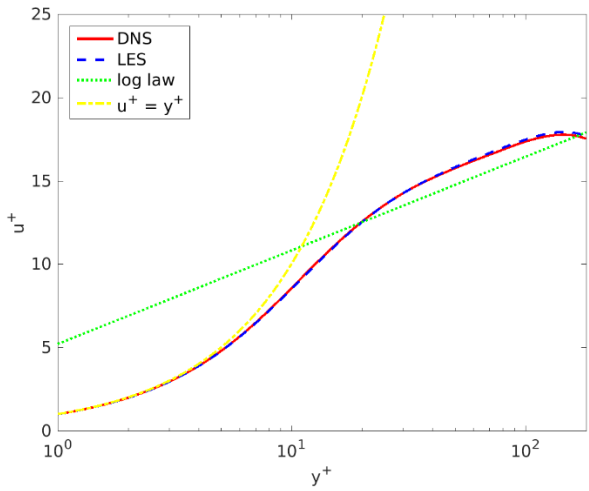
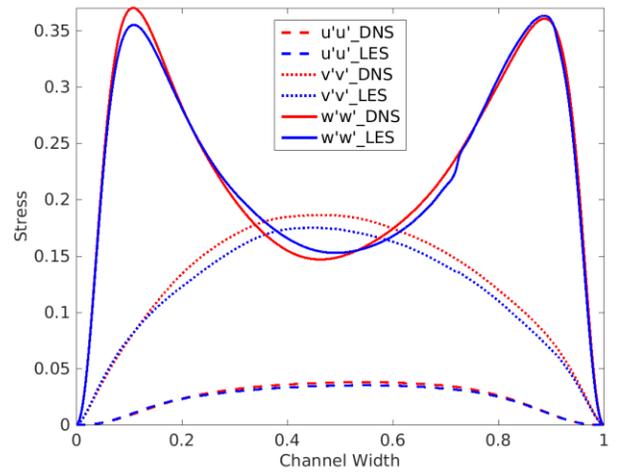

**FIGURE 6b:** NON-DIMENSIONAL VELOCITY PROFILES ACROSS THE CHANNEL HALF-WIDTHS FOR LINES 4 (TOP) THROUGH 6 (BOTTOM)

**FIGURE 7a:** REYNOLDS NORMAL STRESSES FOR LINES 1 (TOP) THROUGH 3 (BOTTOM)

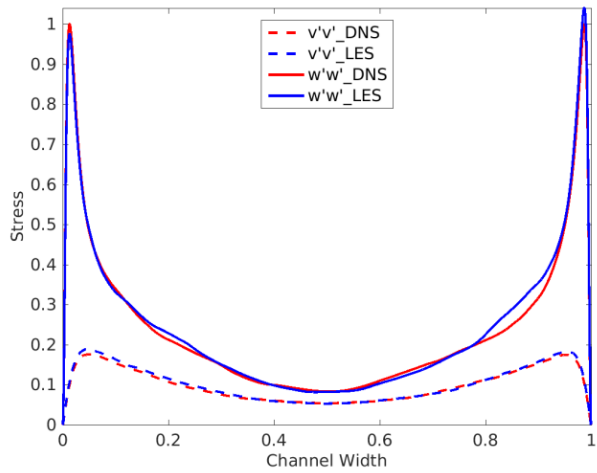
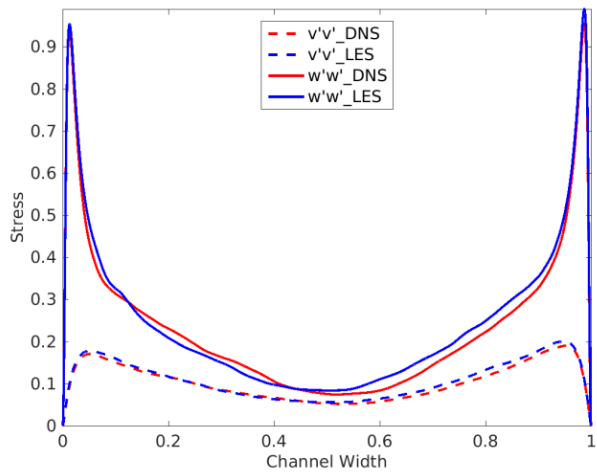
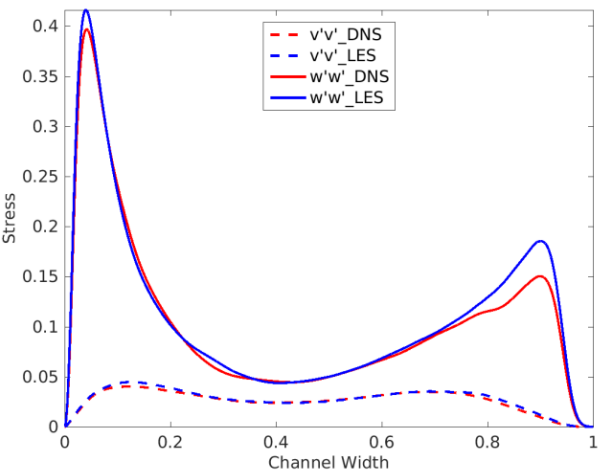

**FIGURE 7b:** REYNOLDS NORMAL STRESSES FOR LINES 4 (TOP) THROUGH 6 (BOTTOM)

## 3.3 Anisotropy Invariants

Further details of the flow field are visualized by plotting functions of the anisotropy tensor invariants. This is done here in the 2D (i.e. streamwise-averaged) domain by use of the componentality contours as presented in Emory and Iaccarino [13]. A barycentric triangle mapping is used to assign an RGB triplet to each of the corners. Red corresponds to one-component turbulence, while green is two-component and blue is three-component. This allows for a detailed spatial representation of the characteristics of the turbulence.

It is clearly shown that the turbulence is primarily one-component in the boundary layers near walls and three-component (i.e. roughly isotropic) far from walls, as in canonical channel flow. However, in the edge subchannels more complex behavior is seen. Notably, there is a shift toward two-component turbulence which is not seen in the interior channels. This behavior is consistent with a gap vortex street as indicated by Merzari et al. [15] and discussed further in Kraus et al. [9].

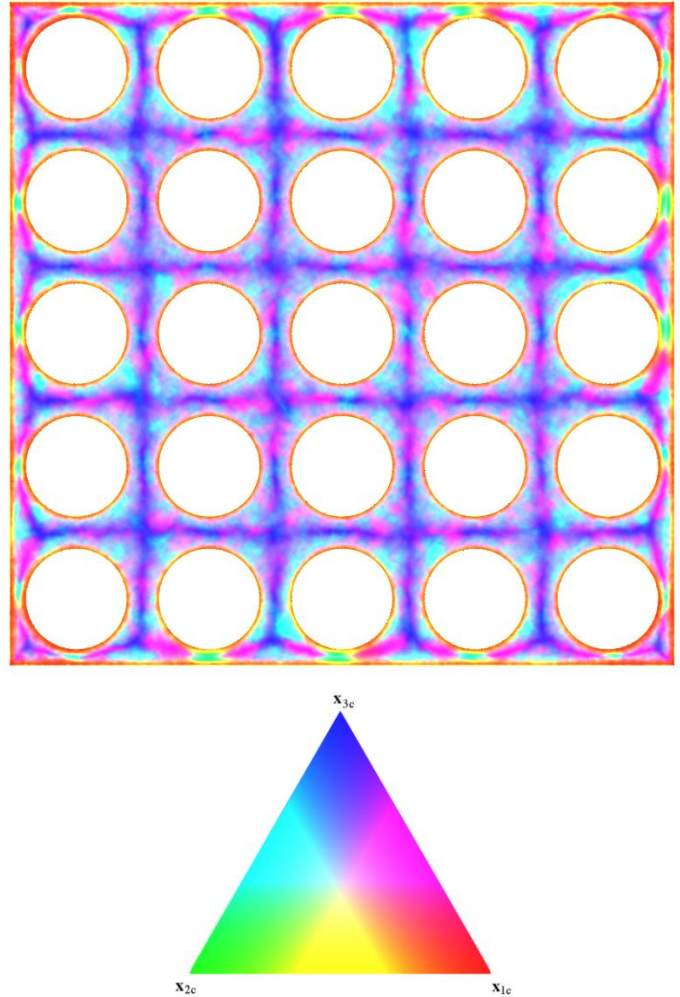

**FIGURE 8:** BARYCENTRIC MAP COMPONENTALITY CONTOURS

## 3.4 Turbulent Kinetic Energy Budget

Analysis of the turbulent kinetic energy budget provides insight into the details of the production, dissipation, and redistribution of turbulence. TKE is transported in virtually every turbulence model, and presentation of the budgets from DNS can provide reference data for assessing the various modeling coefficients. For the case of incompressible flow with constant density and viscosity as investigated here, the TKE budget can be written as:

$$\frac{\partial k}{\partial t} + \overline{u_j}\frac{\partial k}{\partial x_j} = -\frac{1}{\rho}\frac{\partial \overline{u_i' p'}}{\partial x_i} - \frac{1}{2}\frac{\partial \overline{u_j' u_j' u_i'}}{\partial x_i} + \nu \frac{\partial^2 k}{\partial x_j^2} - \overline{u_i' u_j'}\frac{\partial \overline{u_i}}{\partial x_j} - \nu \overline{\frac{\partial u_i'}{\partial x_j}\frac{\partial u_i'}{\partial x_j}} \quad (4)$$

where the terms on the right-hand side represent, respectively, pressure diffusion, turbulent transport, viscous diffusion, production, and dissipation. Outside of the boundary layer, production and dissipation are typically the dominant terms.

Fig. 9 provides TKE budgets for lines 1, 3, and 5. The shapes of the production are all similar, peaking at $y^+ \approx 11$, and are highest in the wide gaps. For lines 1 and 5, production and dissipation are the dominant terms at larger $y^+$ and the redistribution terms are small. For the edge subchannel gap, however, the behavior is notably different. The production is roughly zero at the center of the subchannel. Turbulent transport is the dominant source of TKE in this region, along with some increased contributions from pressure diffusion as well. This is also consistent with previous observations of Merzari et al. [15] and Lai et al. [16], where dominance of the transport term was similarly observed.

## 4. CONCLUSION

Results from the DNS of a square 5x5 rod bundle show the general complexity of flow phenomena in this class of flows. The impacts of non-homogeneous cross-section were most evident in the edge subchannels, where a potential gap vortex street leads to changes in the nature of the localized turbulence. The authors expect this DNS will provide useful data for assessing and developing various reduced-resolution modeling approaches.

Similar work can be performed to provide even more beneficial results. For example, other heterogeneities can impact the flow beyond just the outer wall. These include control rod guide tubes. Thus a case with inhomogeneity in the central bundle region would provide an interesting comparison case to the current work, where the interior subchannels were found to behave similarly. The impacts of heat transfer, including computation of the temperature variance budget, would also be impactful for assessing turbulent heat flux modeling approaches.

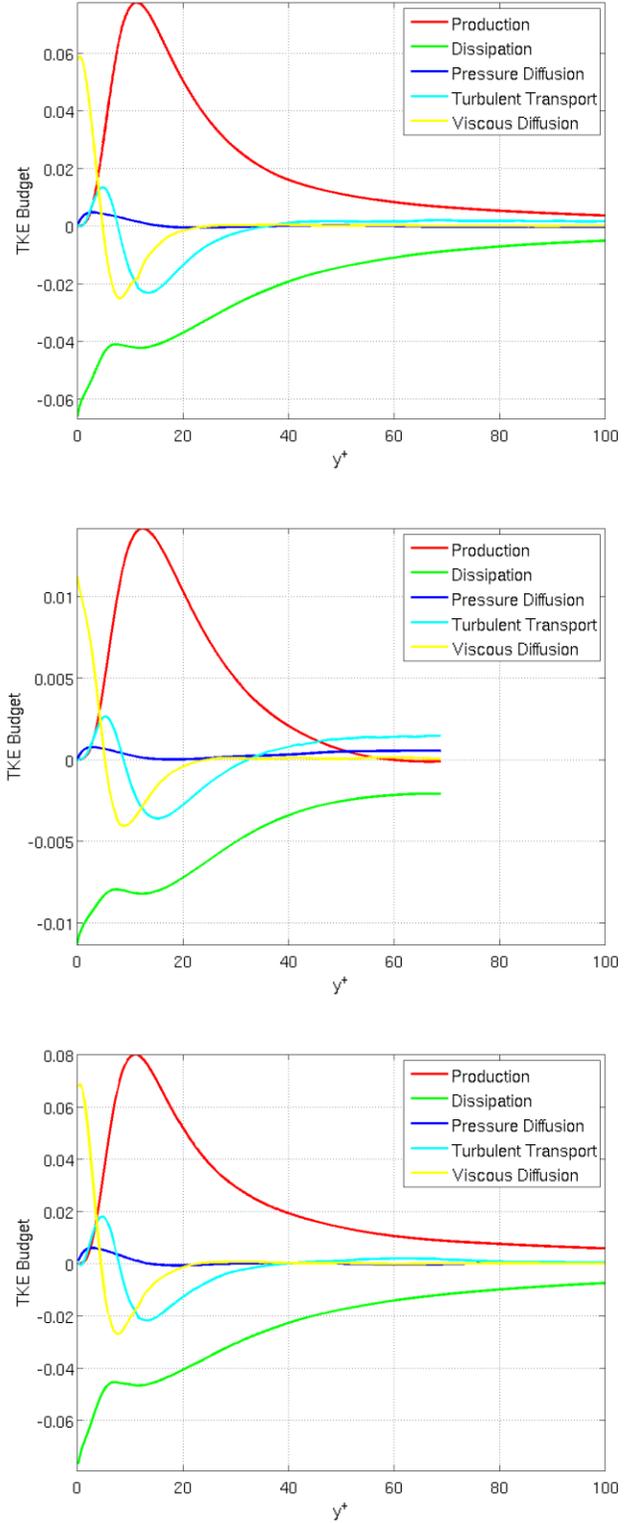

**FIGURE 9:** TURBULENT KINETIC ENERGY BUDGETS FOR LINES 1, 3, AND 5

**ACKNOWLEDGEMENTS**

This research used resources of the Argonne Leadership Computing Facility, which is a DOE Office of Science User Facility supported under Contract DE-AC02-06CH11357. We also gratefully acknowledge the computing resources provided on Bebop, a high-performance computing cluster operated by the Laboratory Computing Resource Center at Argonne National Laboratory.

**REFERENCES**

[1] Nek5000, 2017, "Nek5000 17.0 Documentation," UChicago Argonne, Chicago, IL, https://nek5000.github.io/NekDoc

[2] A. T. Patera, "A Spectral Element Method for Fluid Dynamics: Laminar Flow in a Channel Expansion," J. Comput. Phys., 54(3), pp. 468–488 (1984).

[3] P. F. Fischer, "An Overlapping Schwartz Method for Spectral Element Solution of the Incompressible Navier-Stokes Equations," *J. Comp. Phys.*, **133**(1), pp. 84-101 (1997).

[4] J. Walker, E. Merzari, A. Obabko, P. Fischer, A. Siegel, "Accurate Prediction of the Wall Shear Stress in Rod Bundles with the Spectral Element Method at High Reynolds Number," *Int. J. Heat Fluid Flow*, **50**, pp. 287-299 (2014).

[5] A. Obabko, P. Fischer, O. Marin, E. Merzari, D. Pointer, "Verification and Validation of Nek5000 for T-Junction, Matis, SIBERIA, and MAX experiments," *Proc. 16th Intl. Topical Meeting on Nuclear Reactor Thermal Hydraulics (NURETH-16)*, Chicago, IL, USA, Aug. 30-Sep. 4, 2015.

[6] A. V. Obabko, et al., "Cross-Verification and Validation Simulations of Matis Benchmark," Argonne National Laboratory Report, ANL/TM-341, 2015.

[7] V. Makarashvili, E. Merzari, A. Obabko, H. Yuan, K. Karazis, "Nek5000 Simulations on Turbulent Coolant Flow in a Fuel Assembly Experiment," Argonne National Laboratory Technical Report, ANL/MCS-TM-376, 2018.

[8] B. Mikuz and I. Tiselj, "Wall-resolved Large Eddy Simulation in Grid-free 5 × 5 Rod Bundle of MATiS-H Experiment," *Nucl. Eng. Des.*, **298**, pp. 64-77 (2016).

[9] A. R. Kraus, E. Merzari, T. Norddine, O. Marin, and S. Benhamadouche, "Towards Direct Numerical Simulation of a 5x5 Rod Bundle," International Topical Meeting on Advances in Thermal Hydraulics (ATH 2020), Oct. 20-23, 2020, Palaiseau, France.

[10] STAR-CCM+, © CD-adapco, LTD (A Siemens Company). Available: https://mdx.plm.automation.siemens.com/star-ccm-plus

[11] R. Manceau, "Recent Progress in the Development of the Elliptic Blending Reynolds-stress Model," *Int. J. Heat Fluid Flow*, **51**, pp. 195-220 (2015).

[12] J. Kim, P. Moin, and R. Moser, "Turbulence Statistics in Fully Developed Channel Flow at Low Reynolds Number," *J. Fluid Mech.*, **177**, pp. 133-166 (1987).

[13] M. Emory and G. Iaccarino, "Visualizing Turbulence Anisotropy in the Spatial Domain with Componentality Contours," in *Annual Research Briefs*, Center for Turbulence Research, Stanford University, pp. 123–139 (2014).

[14] S. Tavoularis, "Rod Bundle Vortex Networks, Gap Vortex Streets, and Gap Instability: A Nomenclature and Some Comments on Available Methodologies," *Nucl. Eng. Des.*, **241**(7), pp. 2624–2626 (2011).

[15] E. Merzari and H. Ninokata, "Anisotropic Turbulence and Coherent Structures in Eccentric Annular Channels," *Flow, Turbulence and Combustion*, **82**(1), pp. 93-120 (2009).

[16] J. K. Lai, G. Busco, E. Merzari, Y. A. Hassan, "Direct Numerical Simulation of the Flow in a Bare Rod Bundle at Different Prandtl Numbers." *Journal of Heat Transfer* **141**(12), (2019).